\documentclass[12pt]{article} 

\usepackage{amsmath}
\usepackage{bbold}
\usepackage{float}
\usepackage[top=0.84in, bottom=0.78in, left=0.67in, right=0.67in]{geometry}

\usepackage{graphicx}
\usepackage{cite}
\usepackage{color}
\usepackage{soul}

\newcommand{\braket}[2]{\langle#1|#2\rangle} 
\newcommand{\ket}[1]{|#1\rangle}                     
\newcommand{\bra}[1]{\langle #1|}

\newcommand{\etal}{\textit{et al.}}                  
\title{Direct measurement of large-scale quantum states}

\author{Eliot Bolduc$^{1,*}$, Genevieve Gariepy$^{1}$ and Jonathan Leach$^{1}$\\
\normalsize{$^{1}$Institute of Photonics and Quantum Sciences, Heriot-Watt University,} \\\normalsize{David Brewster Building, Edinburgh, EH14 4AS, UK} \\\small{$^*$eb154@hw.ac.uk}}

\date{}

\begin{document}

\maketitle

\textbf{In quantum mechanics, predictions are made by way of calculating expectation values of observables, which take the form of Hermitian operators.  It is far less common to exploit non-Hermitian operators to perform measurements. Here, we show that the expectation values of a particular set of non-Hermitian matrices, which we call column operators, directly yield the complex coefficients of a quantum state vector.   We provide a definition of the state vector in terms of measurable quantities by decomposing the column operators into observables. The technique we propose renders  very-large-scale quantum states significantly more accessible in the laboratory, as we demonstrate by experimentally characterising a 100 000-dimensional entangled state. This represents an improvement of two orders of magnitude with respect to previous characterisations of discrete entangled states. Furthermore, in numerical studies, we consider mixed quantum states and show that for purities greater that 0.81, we can reliably extract the most significant eigenvector of the density matrix with a probability greater than 99\%.  We anticipate that our method will prove to be a useful asset in the quest for understanding and manipulating large-scale quantum systems.}

\paragraph{} One of the current challenges in the field of computing is harnessing the potential processing power provided by quantum devices that exploit entanglement.  Experimental research aimed at overcoming this challenge is  driven by the production, control and detection of larger and larger entangled quantum states  \cite{Monz:2011,Wong:2012,Yokoyama:2013,Krenn:2014}. However, the task of characterising these entangled states quickly becomes intractable as the number of parameters that define a many-body system scales exponentially with the system size.  To keep up with the ever-growing quantum state dimensionality, much effort is put into developing efficient characterisation methods  \cite{Smith:2005,Banaszek:2013,Flammia:2005,Cramer:2010,Bogdanov:2010,Toth:2010,Gross:2010,Mahler:2013,Schwemmer:2014,Shabani:2011,Teo:2013,Tonolini:2014,Lloyd:2014,Ferrie:2014,Lundeen:2011}.

\paragraph{}  Quantum state tomography is the process of retrieving the values that define a quantum system. The process typically involves two steps: i) gathering an informationally complete set of data and ii) finding the quantum state most consistent with the data set using post-measurement processing such as the maximum likelihood estimation algorithm \cite{Banaszek:1999}. Many efficient tomographic methods capitalize on the first step by making simplifying assumptions about the state\cite{Flammia:2005,Cramer:2010,Toth:2010,Schwemmer:2014,Gross:2010,Shabani:2011,Tonolini:2014,Lloyd:2014,Lundeen:2011}, thus reducing the number of measurements required to uniquely identify it. In particular, tomography via compressed sensing allows one to efficiently reconstruct quantum states based on the fact that  low-rank density matrices, i.e.~quasi-pure states,  are sparse in a particular basis  \cite{Gross:2010,Liu:2012,Schwemmer:2014,Tonolini:2014}. Compared to assumption-free tomography, compressive sensing provides a square-root improvement on the required number of measurements \cite{Banaszek:2013}. This improvement enabled the reconstruction of the density matrices of a 6-qubit state \cite{Schwemmer:2014} and a (17$\times$17)-dimensional state \cite{Tonolini:2014}, the largest phase-and-amplitude measurement of an entangled state reported to date. Although compressed sensing does not make use of maximum likelihood estimation, it does require non-trivial post-measurement processing.

 \paragraph{} Recently, Lundeen \etal~reported on the direct measurement of a wavefunction using a method that, for the first time, required no involved post-measurement processing \cite{Lundeen:2011}. Their method is based on weak measurements, whereby one weakly couples a quantum system to a pointer state and subsequently performs a few standard strong measurements on the pointer state. The outcome of a weak measurement is known as the ``weak value", and in the conditions exposed in Ref. \cite{Lundeen:2011} the weak value is proportional to a given state vector coefficient. The method of Lundeed \etal~can be used in combination with the assumption that the quantum state at hand is pure, providing the same square-root improvement as compressed sensing. Variations on the original scheme allow measurements of mixed states and increased detection efficiency \cite{Bamber:2014,Salvail:2013,Wu:2013gb}.

 \paragraph{}  An important contribution of the work by Lundeen \etal~was to link the state vector elements to the expectation value of weak measurements. We take a different approach, and point out that the enabling feature that allows access to the complex state vector is not weak measurement but the use of particular non-Hermitian operators. Although weak measurements provide a way to decompose these non-Hermitian operators, it is not the only suitable approach. Moreover, the introduction of weak values in the measurement procedure adds complexity to the experiment and the formalism that links weak values to measurement outcomes involves an approximation that breaks down in a variety of circumstances \cite{Duck:1989,Salvail:2013,Malik:2013}.

 \paragraph{}  In this paper, we propose an alternative approach to the direct measurement of quantum states that is exact in the case of pure states, proves to be reliable in the presence of noise, and is consistent with results obtained with well-established tomographic techniques. The key principle of our formalism is to decompose the particular non-Hermitian matrices that yield the complex state vector coefficients using only observables.  Our method therefore only requires strong measurements, as in standard tomography, while maintaining the directness of weak-value-assisted tomography.  The simplicity in both the experimental procedure and post-measurement processing renders our method ideally suited for the characterisation of  large-scale systems, which can be high-dimensional, many-body or both. We begin by developing the theory on which our method is based and then demonstrate the potential of this scheme by experimentally retrieving the complex coefficients of a (341$\times$341)-dimensional entangled state.

 \paragraph{} Consider a quantum system in a $d$-dimensional Hilbert space, whose state vector
\begin{equation}\label{eq:psi}
\ket{\Psi}=\sum_{j=0}^{d-1} c_j \ket{j}
\end{equation}
is expanded in the basis $\{ \ket{j}\}$ and where $c_j$ are unknown complex expansion coefficients. In order to retrieve these coefficients, we introduce the column operators $\widehat{C}_j=\ket{a}\bra{j}$, where $\ket{a}$ is an arbitrary reference vector. Each column operator has an expectation value
\begin{equation}\label{eq:column}
 \langle{\widehat{C}_j}\rangle=\braket{\Psi}{a}{c_j},
 \end{equation}
which is proportional to a complex state vector expansion coefficient. Since the value of ${\braket{\Psi}{a}}$  is  independent of $j$,  we can express the state vector in terms of the column operators up to a phase factor:
\begin{equation}\label{eq:master}
\ket{\Psi}=\frac{\text{e}^{i\phi}}{\nu}\sum_{j=0}^{d-1} \langle\widehat{C}_j\rangle\ket{j},
\end{equation}
where $\nu = |{\braket{\Psi}{a}}|$ is a normalization constant. We can ignore the phase factor $\text{e}^{i\phi}$ since it bears no physical significance.

\paragraph{} Most column operators $\widehat{C}_j$ are not Hermitian matrices and are thus not observables. To overcome this apparent constraint,  we recognize that any non-Hermitian matrix can be constructed from a complex-weighted sum of Hermitian matrices. Hence, the crucial step to our method is to construct the column operators in terms of measurable quantities: $\widehat{C}_j= \sum_q {w}_{jq} {\widehat{\mathcal O}_{jq}}$, where ${w}_{jq}$ are complex weights and ${\widehat{\mathcal O}_{jq}}$ are observables. As a result, this allows us to retrieve any  state vector element with a complex-weighted sum of measurement outcomes:
\begin{equation}\label{eq:coef}
c_j=\frac{1}{\nu}\sum_{q} {w}_{jq} \langle\widehat{\mathcal{O}}_{jq}\rangle.
\end{equation}
Equation \ref{eq:coef} is an exact definition of the pure state vector that is provided in terms of measurable quantities. The above formalism readily applies to a general class of quantum states, including high-dimensional and many-body systems.

\paragraph{} 
As an example, consider the case of a qubit $\ket{\Psi}=c_0\ket{0}+c_1\ket{1}$ with $\ket{a}=\ket{0}$ as the reference vector.  The first column operator $\widehat{C}_0$ is Hermitian and given by the projector  $\ket{0}\bra{0}$. The second column operator $\widehat{C}_1=\ket{0}\bra{1}$ is not Hermitian but can be constructed a number of ways. The first  construction -- which, as pointed out earlier, is a key part of the weak value formalism --  is the complex-weighted sum of Pauli matrices: $\widehat{C}_1=(\hat\sigma_x+i\hat\sigma_y)/2$, a decomposition that requires two observables, each of which is made of two projectors or eigenvectors. A second decomposition requiring only three projectors is given by 
\begin{equation}\label{eq:tetrarec}
 \widehat{C}_1= \sum_{q=0}^2 \frac{2}{3}\text{e}^{i 2\pi q /3} \ket{s_{q}}\bra{s_{q}},
\end{equation}
where $\ket{s_{q}}=(\ket{0}+\text{e}^{i 4 \pi q/3 } \ket{1})/\sqrt{2}$ are the states onto which the  observables $\widehat{\mathcal{O}}_{1q}$  project. In both cases, the qubit state vector is exactly given by $\ket{\Psi}=(\langle\widehat{C}_0\rangle \ket{0} + \langle\widehat{C}_1\rangle\ket{1})/\langle\widehat{C}_0\rangle^\frac{1}{2}$.

\paragraph{} 
To demonstrate the power and scalability of our scheme, we apply it to the measurement of a state entangled in greater than 100 000 dimensions. We provide a complete characterisation of the spatially entangled two-photon field produced through spontaneous parametric downconversion (SPDC). In general, SPDC can give rise to spatial and frequency correlations between two photons \cite{Miatto:2011,Dada:2011,Agnew:2011,Leach:2012,Salakhutdinov:2012,Tasca:2012,Geelen:2013,Krenn:2014,Osorio:2008,Mosley:2008,Osorio:2013}. The purity of the spatial part of the full state can only be guaranteed if the two types of correlations are completely decoupled, which can be achieved in the collinear regime \cite{Osorio:2008} -- see Supplementary Information section A for a theoretical estimation of our system purity. The consequences of applying our scheme to a quantum state with non-unit purity, which is always the case in the presence of noise, will be discussed below.

\begin{figure}
	\raggedright
	\includegraphics{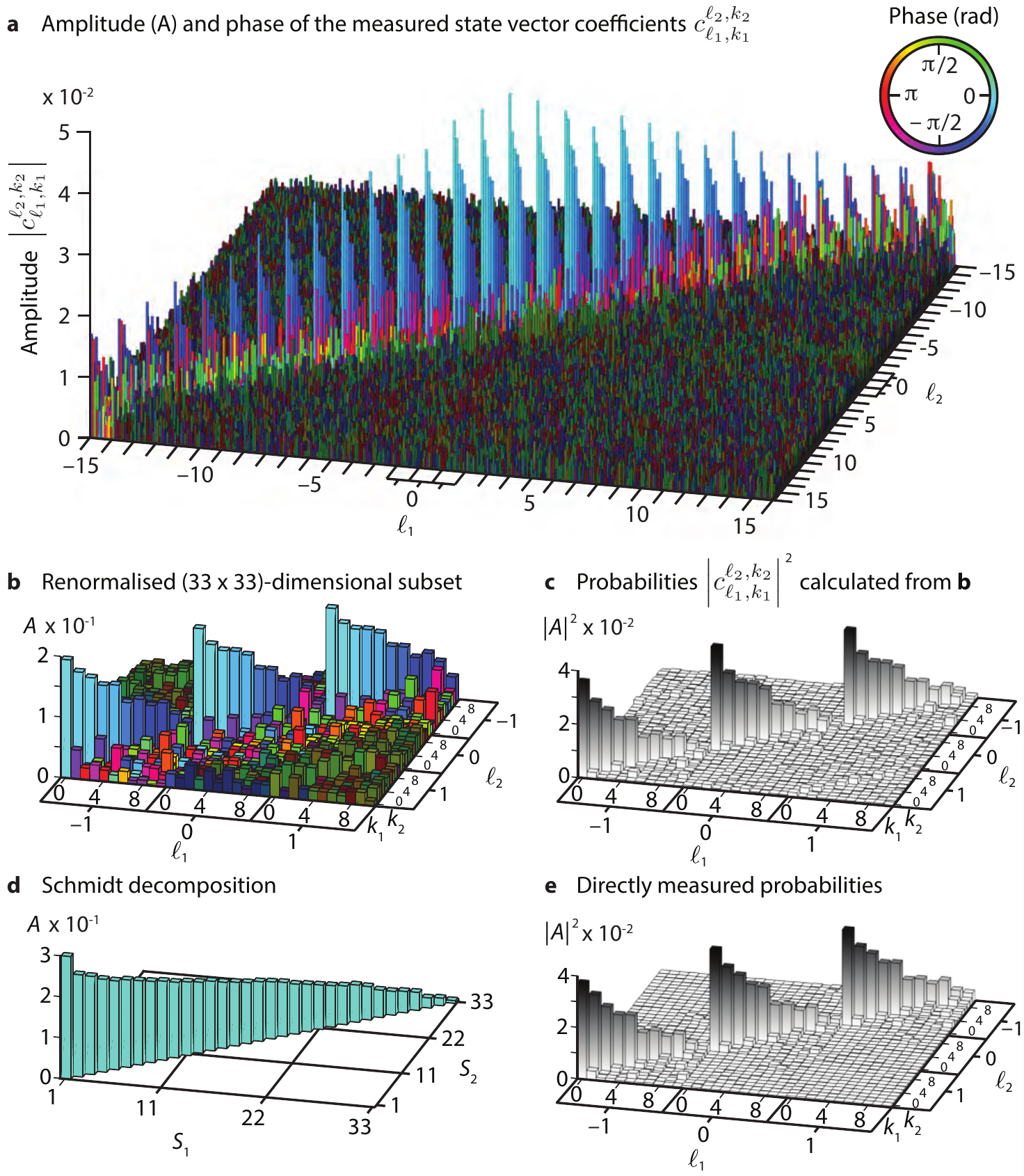}
	\caption{\small{\textbf{Measured and calculated properties of the two-photon state.}  \textbf{a}   We illustrate the complex state vector coefficients in matrix format. This representation is similar to that of an optical transfer matrix, where the lateral axes correspond to input and output modes. Here, each lateral axis corresponds to the spatial state of one photon of the pair. The OAM values $\ell$ range from -15 to 15. For a fixed OAM value, the radial index $k$ ranges from 0 to 10, thus the combined state has $341\times341$ dimensions.  The amplitude and phase values of a  coefficient are given by the height and color of a bar, respectively. For clarity, we darken the off-diagonal part. The small subset \textbf{b} of the state shows the phase gradient across the diagonal elements, which is typical of a Gouy phase shift, a common property of light passing through focus \cite{Boyd79}. The corresponding calculated probability matrix \textbf{c}  is consistent with the directly measured probabilities \textbf{e}. Finally, we calculate the Schmidt decomposition \textbf{d} of subset \textbf{b}, which gives the joint basis in which the subset can be expressed with the lowest number of modes. The indices $S_1$ and $S_2$ correspond to the states of each photon in this basis. }}
	\label{fig:Walsh}
\end{figure}

\paragraph{}  
We express the spatial part of the entangled state in a discrete cylindrical basis of transverse spatial modes. The azimuthal part of the modes is given by $\text{e}^{i\ell\phi}$, where $\ell$ is an integer between $-\infty$ and $\infty$ and $\phi$ is the azimuthal angle. This type of phase profile is known to carry $\ell$ units of orbital angular momentum (OAM). We decompose the radial part of the field with the recently introduced Walsh modes, labelled by the integer $k$ ranging from 0 to $\infty$  \cite{Geelen:2013}. The Walsh modes all have the same Gaussian amplitude envelope, but different $\pi$-steps radial phase profiles. Combining the OAM modes with the Walsh modes yields a complete basis for coherent two-dimensional images. To perform the characterisation of the two-photon spatial field, we consider 31 OAM modes and 11 Walsh modes for each photon. The state vector thus takes the form
\begin{equation}\label{eq:c}
\ket{\Phi} = \sum_{\ell_1=-15}^{15}\sum_{k_1=0}^{10} \sum_{\ell_2=15}^{-15}\sum_{k_2=0}^{10} c_{\ell_1,k_1}^{\ell_2,k_2} \ket{\ell_1,k_1}\ket{\ell_2,k_2}.
\end{equation}
Using the column-operator decomposition described in the Methods section, we sequentially measure all 116 281 coefficients $c_{\ell_1,k_1}^{\ell_2,k_2}$, which are shown in figure \ref{fig:Walsh}a and \ref{fig:Walsh}b. The total Hilbert space dimensionality of this measured state is more than two orders of magnitude larger than any previously reported amplitude-and-phase-characterised discrete entangled state \cite{Tonolini:2014}.  As a simple verification of the accuracy of our method, we calculate the probabilities associated with each joint mode via the Born rule, $|c_{\ell_1,k_1}^{\ell_2,k_2}|^2$, as shown in figure \ref{fig:Walsh}c. This result is consistent with the directly measured correlation matrix shown in figure \ref{fig:Walsh}e, showing that we retrieve the correct magnitude of the amplitudes.


\paragraph{} To rigorously assess the validity of the directly measured complex quantum state $\ket{\psi}$, i.e.~both the amplitudes and the phases, we compare it to the results obtained through full tomography (i.e. assumption-free tomography).  As full tomography cannot be performed on a (341$\times$341)-dimensional entangled state in a reasonable time, we characterise a ($5\times 5$)-dimensional subset of the SPDC two-photon state. We perform the comparison in a basis of various OAM modes ($\ell_1 \in \{1,-1,2,-2,3 \}$, $\ell_2 \in \{1,-1,2,-2,-3 \}$ ) and a fixed radial Walsh mode ($k_1=k_2=0$). The total number of unknown parameters in the corresponding density matrix is equal to $624$. After performing the direct measurement procedure in this basis, we record 8000 random projective measurements that we break into 8 sets of 1000. For each set, we recover a density matrix $\rho_{\textrm{exp}}$ and calculate its purity and the fidelity with the directly measured state $\ket{\psi}$; fidelity is defined as $\sqrt{\bra{\psi}\rho_{\textrm{exp}}\ket{\psi}}$. On average, the purity calculation yields ($0.96\pm0.02$), and the fidelity gives ($0.985\pm0.004$), where the uncertainties correspond to one standard deviation. After reconstruction of a density matrix, we find that the average error between the measured count rates and the count rates predicted by the density matrix is 5.5\%. This can be explained by shot noise, the pixelated nature of the SLM, and the finite aperture of the optical elements. While we expect unit purity, the 5\% noise level accounts for the discrepancy with the measured value.  

\paragraph{} The extremely high fidelity between the tomography results $\rho_{\textrm{exp}}$ and the directly measured state $\ket{\psi}$ indicates the validity of our approach for quantum state measurements applied to near pure states.  To evaluate our method in the context of mixed states, we perform a series of numerical simulations where we vary the rank, purity, and dimension of an unknown state $\rho_{\textrm{sim}}$, where no sources of noise are added to the simulated measurement outcomes.  We apply our direct measurement procedure to these states and calculate the fidelity $|\braket{\psi}{\psi_{\textrm{sim}}}|$, where $\ket{\psi_{\textrm{sim}}}$ is the eigenvector of $\rho_{\textrm{sim}}$ with the largest eigenvalue.  For initial states $\rho_{\textrm{sim}}$ with purity greater than $0.81$, we measure a fidelity greater than $0.99$ in at least 99\% of the cases.  The dependency of this result on the dimensionality of the state is negligible. This result indicates that our direct method is able to extract the primary eigenvector of a density matrix, even for a partially mixed state. Full details of this analysis and the density matrix reconstruction are presented in the Supplementary Information.

 \paragraph{}  Knowledge of the amplitude and phase of the state vector elements allows us to perform otherwise inaccessible calculations. As an example, we perform a calculation of the Schmidt decomposition \cite{Ekert:1995}. This is equivalent to the singular value decomposition for the case of optical transfer matrices. The Schmidt decomposition yields a new joint basis in which the photons are perfectly correlated and where the joint modes have equal phases, as shown in figure \ref{fig:Walsh}d. When the Schmidt decomposition is applied to the entire state, we calculate a number of Schmidt modes equal to 142; this represents the effective number of independent joint modes contained within the state (the maximum for a (341$\times$341)-dimensional state being 341). The Schmidt decomposed two-photon field is a good candidate for the violation of very-high-dimensional Bell inequalities \cite{Dada:2011}.   Further details on the Schmidt decomposition can be found in the Supplementary Information.

 \paragraph{} There are a number of approaches to reducing the necessary cost and effort for measuring large-scale quantum states.  These include, but are not limited to, developing technologies for mode sorting \cite{Berkhout:2010} and arbitrary unitary transformations \cite{Morizur:2010,Miller:2013}, reducing the required number of measurement settings, and circumventing the requirement for reconstruction procedures.  It is clear that there is significant interplay between each of these approaches.  The theoretical implementation of an approach that combines the principles of our work with generalised measurements, such as POVMs (positive operator value measures), is considered in the Supplementary Information. The ability to use POVMs in the laboratory relies on the aforementioned technologies.  Access to these types of technologies would reduce the overall number of measurement settings to uniquely recover a quantum state.  However, such a system requires arbitrary unitary transformations for spatial states, which is in itself an active area of research  \cite{Berkhout:2010,Morizur:2010,Miller:2013}.  Given the limitations of mode sorters for very large dimensions, and the practical nature of projective measurements, our scheme provides a simple and elegant method for the characterisation of large-scale quantum states.


 \paragraph{} Our scheme allows direct access to the complex coefficients that define large-scale quantum states. The main result of our work is a novel method for retrieving a state vector coefficient with a complex-weighted sum of strong measurement outcomes.  One challenge in reconstructing a quantum state from measurement outcomes lies in data processing; our scheme trades the difficulty of data processing for theoretical analysis prior to the experiment, that is, finding the measurements one has to perform. We anticipate that our work will have an impact on a number of disciplines, for example, quantum parameter estimation, measurement in quantum computing, quantum information and metrology. \\

\begin{figure}
	\raggedright
	\includegraphics{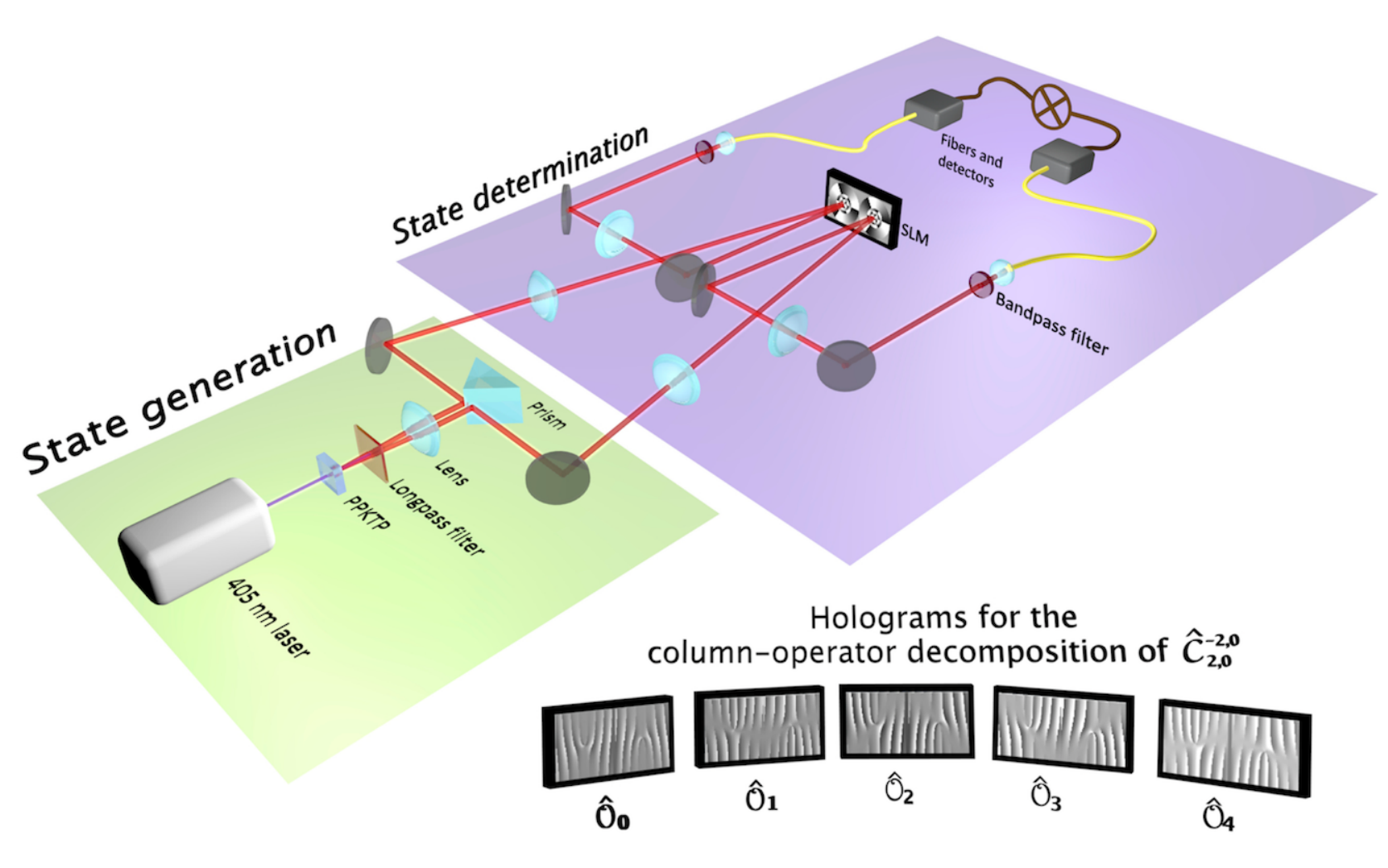}
	\caption{\textbf{Generation and characterisation of a two-photon field.} The entangled state is produced via SPDC in a PPKTP crystal and spatially separated by a prism. For the state determination stage, the crystal plane is imaged onto a spatial light modulator (SLM), which is in turn imaged to the input facet of two single mode fibers. In order to make a given projective measurement, we display the corresponding joint mode on the SLM and measure the coincidence rate between the two single photon avalanche diode detectors. The inset shows the five joint holograms that correspond to the column-operator decomposition of $\widehat{C}_{2,0}^{\hspace{1pt}\text{-}2,0}$. The state vector coefficient $c_{2,0}^{\text{-}2,0}$ is given by $\frac{4}{5\nu}\sum_q \langle{\mathcal{\widehat{O}}_q}\rangle \text{e}^{i 4\pi q/5}$, where the expectation value of a given observable is proportional to the measured count rate when displaying the corresponding hologram.}
	\label{fig:setup}
\end{figure}

 \noindent \textbf{Methods}\\
 \textbf{Experiment}
 The  two-photon field is generated via SPDC with a 405-nm laser diode pumping a 1-mm-long periodically-poled KTP (PPKTP) crystal with 50 mW of power. The experimental setup is shown in figure \ref{fig:setup}. We separate the two photons  with a right angle prism and image the plane of the crystal to a Holoeye spatial light modulator (SLM) with a magnification of -10. We  simultaneously display two holograms, one on each side of the SLM, to control the amplitude and phase profiles of the two photons independently. In order to make projective measurements of superposition modes, we make use of intensity masking \cite{Bolduc:2013}. We image the plane of the SLM with a magnification of $-1/2500$ to two single mode fibers. The combination of the SLM and singles mode fibers allows us to make arbitrary projective measurements. All measurements are performed in coincidence with two single photon avalanche detectors, with a timing window of 25 ns, an integration time of 1 s for modes outside the diagonal and 20 s for the diagonal elements ($\ell_1=-\ell_2$ and $k_1=k_2$). We start an automatic alignment procedure with the SLM every four hours to compensate for drift. Including the time it takes to calculate and display a hologram (about one second), the entire experiment takes two weeks; assumption-free tomography would take more than four centuries at the same acquisition rate. We perform no background subtraction and use the fundamental mode ($\ell_1=-\ell_2=k_1=k_2=0$) as the reference vector $\ket{a}$. The count rate of the fundamental mode is approximately 900 coincidences per second and varies by 10\% over 24-hour periods. To correct for long term drift, we normalise each outcome to the count rate of the fundamental mode, which we measure before the measurement of each column operator. In standard tomography, the calculation of error bounds on the measured state is not a straightforward task \cite{Christandl:2012}. Here, we can calculate the error bound on a given  coefficient with a weighted sum of the detector counts used to retrieve it. For a given state vector coefficient, the errors on the amplitude $|c_j|$ and phase arg$(c_j)$ are both inversely proportional to the overlap $\nu$ of the reference vector with the quantum state. In order to minimize the errors, it is important to choose a reference vector that has a high probability of occurrence within the state -- the fundamental mode is the most probable one in our case. \\

\noindent \textbf{Two-body column-operator decomposition} In order to decompose a given state vector coefficient $c_{\ell_1,k_1}^{\ell_2,k_2}$ into a set of measurement  outcomes, we need to find a projector decomposition of  the corresponding column operator $\widehat{C}_{\ell_1,k_1}^{\ell_2,k_2}=\ket{0,0}\bra{\ell_1,k_1}\otimes\ket{0,0}\bra{\ell_2,k_2}$, as in equation \ref{eq:coef}. We numerically find this column-operator decomposition, i.e. the  complex weights ${w}_q$ and the observables $\widehat{\mathcal{O}}_q$, using the differential evolution algorithm (see Supplementary Information part D). By inspection, we find that the corresponding analytical form of the state vector coefficients is given by 
\begin{equation}\label{eq:coef5}
\widehat{C}_{\ell_1,k_1}^{\ell_2,k_2}=\frac{1}{\nu} \sum_{q=0}^{4} \frac{4}{5} \text{e}^{i 2\pi q /5}\ket{s_{1,q}}\bra{s_{1,q}}\otimes\ket{s_{2,q}}\bra{s_{2,q}}, 
\end{equation}
where $\sqrt{2}\ket{s_{m,q}}=\ket{0,0}+\text{e}^{i 4 \pi q/5} \ket{\ell_m,k_m} $ with $m=\{1,2 \}$, and $\nu=|\braket{\Psi}{0,0}|$ is a normalisation constant. This decomposition is only valid when the state of any  photon is different from the reference vector, i.e. $\ket{\ell_m,k_m}\neq\ket{0,0}$. Each coefficient measured with the above column-operator decomposition requires five projective measurements, thus explaining the $5D^2$ scaling, where $D$ is the Hilbert space dimensionality of a single particle.   The protocol scales much more favorably than assumption-free tomography, which requires $D^4$ projections.

\paragraph{} Here, we briefly explain our protocol for measuring the entire SPDC state vector. We measure more than 99\% of the coefficients using the decomposition of equation \ref{eq:coef5}.  The remaining column operators are the special cases $\ket{0,0}\bra{\ell_1,k_1}\otimes\ket{0,0}\bra{0,0}$ and $\ket{0,0}\bra{0,0}\otimes\ket{0,0}\bra{\ell_2,k_2}$, which respectively correspond to a row and a column of the result shown in figure \ref{fig:Walsh}a. These column operators can be decomposed into only three joint local measurements using the projector $\ket{0,0}\bra{0,0}$ on one system and a column-operator decomposition similar to that of equation \ref{eq:tetrarec} on the other system. Finally, the column operator $\ket{0,0}\bra{0,0}\otimes\ket{0,0}\bra{0,0}$ is a projector, and its expectation value can be measured in a single experimental configuration. \\

\noindent \textbf{Full quantum tomography} We perform full tomography with high count rates in order to achieve high accuracy. We set the magnification between the plane of the SLM and that of the single mode fibers to 1/400. In this condition, we obtain a count rate of approximately 18,000 counts per second for the fundamental mode and integrate over 1 second for each individual projective measurement. The increase in the count rate of the fundamental mode comes at the price of lower count rates for high order modes. Regarding the full tomography measurements, we take an overcomplete set of 1000 random projective measurements in a $(5\times5)$-dimensional space. To minimize high-frequency components on the SLM, we limit the random superpositions to two-dimensional subsets of the state space.


\clearpage

\makeatletter
\setlength{\@fpbot}{1cm}
\makeatother

\makeatletter
\setlength{\@fpbot}{1cm}
\makeatother

 \clearpage

\makeatletter
\setlength{\@fpbot}{1cm}
\makeatother


\begin{thebibliography}{10}
\expandafter\ifx\csname url\endcsname\relax
  \def\url#1{\texttt{#1}}\fi
\expandafter\ifx\csname urlprefix\endcsname\relax\def\urlprefix{URL }\fi
\providecommand{\bibinfo}[2]{#2}
\providecommand{\eprint}[2][]{\url{#2}}


\bibitem{Monz:2011}
\bibinfo{author}{Monz, T.} \emph{et~al.}
\newblock \bibinfo{title}{{14-Qubit Entanglement: Creation and Coherence}}.
\newblock \emph{\bibinfo{journal}{Phys. Rev. Lett.}}
  \textbf{\bibinfo{volume}{106}}, \bibinfo{pages}{130506}
  (\bibinfo{year}{2011}).

\bibitem{Wong:2012}
\bibinfo{author}{Yao, X.~C.} \emph{et~al.}
\newblock \bibinfo{title}{{Observation of eight-photon entanglement}}.
\newblock \emph{\bibinfo{journal}{Nature Photon.}}
  \textbf{\bibinfo{volume}{6}}, \bibinfo{pages}{225--228}
  (\bibinfo{year}{2012}).

\bibitem{Yokoyama:2013}
\bibinfo{author}{Yokoyama, S.} \emph{et~al.}
\newblock \bibinfo{title}{{Ultra-large-scale continuous-variable cluster states
  multiplexed in the time domain}}.
\newblock \emph{\bibinfo{journal}{Nature Photon.}}
  \textbf{\bibinfo{volume}{7}}, \bibinfo{pages}{982--986}
  (\bibinfo{year}{2013}).

\bibitem{Krenn:2014}
\bibinfo{author}{Krenn, M.} \emph{et~al.}
\newblock \bibinfo{title}{{Generation and confirmation of a
  (100x100)-dimensional entangled quantum system}}.
\newblock \emph{\bibinfo{journal}{Proc. Natl Acad. Sci.}}
  \textbf{\bibinfo{volume}{111}}, \bibinfo{pages}{6243--6247}
  (\bibinfo{year}{2014}).





\bibitem{Smith:2005}
\bibinfo{author}{Smith, B.~J.}, \bibinfo{author}{Killett, B.},
  \bibinfo{author}{Raymer, M.~G.}, \bibinfo{author}{Walmsley, I.~A.} \&
  \bibinfo{author}{Banaszek, K.}
\newblock \bibinfo{title}{{Measurement of the transverse spatial quantum state
  of light at the single-photon level}}.
\newblock \bibinfo{type}{Tech. Rep.} (\bibinfo{year}{2005}).

\bibitem{Bogdanov:2010}
\bibinfo{author}{Bogdanov, Y.~I.} \emph{et~al.}
\newblock \bibinfo{title}{{Statistical Estimation of the Efficiency of Quantum
  State Tomography Protocols}}.
\newblock \emph{\bibinfo{journal}{Phys. Rev. Lett.}}
  \textbf{\bibinfo{volume}{105}}, \bibinfo{pages}{010404}
  (\bibinfo{year}{2010}).
  
  

  
  \bibitem{Mahler:2013}
\bibinfo{author}{Mahler, D.~H.} \emph{et~al.}
\newblock \bibinfo{title}{{Adaptive Quantum State Tomography Improves Accuracy
  Quadratically}}.
\newblock \emph{\bibinfo{journal}{Phys. Rev. Lett.}}
  \textbf{\bibinfo{volume}{111}}, \bibinfo{pages}{183601}
  (\bibinfo{year}{2013}).

\bibitem{Teo:2013}
\bibinfo{author}{Teo, Y.~S.}, \bibinfo{author}{{\v{R}}eh{\'a}{\v c}ek, J.} \&
  \bibinfo{author}{Hradil, Z.}
\newblock \bibinfo{title}{{Informationally incomplete quantum tomography}}.
\newblock \emph{\bibinfo{journal}{Quantum Measurements and Quantum Metrology}}
  \textbf{\bibinfo{volume}{1}}, \bibinfo{pages}{57--83}.

\bibitem{Ferrie:2014}
\bibinfo{author}{Ferrie, C.}
\newblock \bibinfo{title}{{Self-Guided Quantum Tomography}}.
\newblock \emph{\bibinfo{journal}{Phys. Rev. Lett.}}
  \textbf{\bibinfo{volume}{113}}, \bibinfo{pages}{190404}
  (\bibinfo{year}{2014}).
  

\bibitem{Banaszek:2013}
\bibinfo{author}{Banaszek, K.}, \bibinfo{author}{Cramer, M.} \&
  \bibinfo{author}{Gross, D.}
\newblock \bibinfo{title}{{Focus on quantum tomography}}.
\newblock \emph{\bibinfo{journal}{New J. Phys.}} \textbf{\bibinfo{volume}{15}},
  \bibinfo{pages}{125020} (\bibinfo{year}{2013}).
  
\bibitem{Shabani:2011}
\bibinfo{author}{Shabani, A.} \emph{et~al.}
\newblock \bibinfo{title}{{Efficient Measurement of Quantum Dynamics via
  Compressive Sensing}}.
\newblock \emph{\bibinfo{journal}{Phys. Rev. Lett.}}
  \textbf{\bibinfo{volume}{106}}, \bibinfo{pages}{100401}
  (\bibinfo{year}{2011}). 
  
\bibitem{Flammia:2005}
\bibinfo{author}{Flammia, S.~T.}, \bibinfo{author}{Silberfarb, A.} \&
  \bibinfo{author}{Caves, C.~M.}
\newblock \bibinfo{title}{{Minimal Informationally Complete Measurements for
  Pure States}}.
\newblock \emph{\bibinfo{journal}{Found Phys}} \textbf{\bibinfo{volume}{35}},
  \bibinfo{pages}{1985--2006} (\bibinfo{year}{2005}).
  
  \bibitem{Cramer:2010}
\bibinfo{author}{Cramer, M.} \emph{et~al.}
\newblock \bibinfo{title}{{Efficient quantum state tomography}}.
\newblock \emph{\bibinfo{journal}{Nat. Commun.}} \textbf{\bibinfo{volume}{1}},
  \bibinfo{pages}{149--7} (\bibinfo{year}{2010}).
  

  
  \bibitem{Toth:2010}
\bibinfo{author}{T{\'o}th, G.} \emph{et~al.}
\newblock \bibinfo{title}{{Permutationally Invariant Quantum Tomography}}.
\newblock \emph{\bibinfo{journal}{Phys. Rev. Lett.}}
  \textbf{\bibinfo{volume}{105}}, \bibinfo{pages}{250403}
  (\bibinfo{year}{2010}).

\bibitem{Gross:2010}
\bibinfo{author}{Gross, D.}, \bibinfo{author}{Liu, Y.-K.},
  \bibinfo{author}{Flammia, S.~T.}, \bibinfo{author}{Becker, S.} \&
  \bibinfo{author}{Eisert, J.}
\newblock \bibinfo{title}{{Quantum State Tomography via Compressed Sensing}}.
\newblock \emph{\bibinfo{journal}{Phys. Rev. Lett.}}
  \textbf{\bibinfo{volume}{105}}, \bibinfo{pages}{150401}
  (\bibinfo{year}{2010}).
 
  


\bibitem{Schwemmer:2014}
\bibinfo{author}{Schwemmer, C.} \emph{et~al.}
\newblock \bibinfo{title}{{Experimental Comparison of Efficient Tomography
  Schemes for a Six-Qubit State}}.
\newblock \emph{\bibinfo{journal}{Phys. Rev. Lett.}}
  \textbf{\bibinfo{volume}{113}}, \bibinfo{pages}{040503}
  (\bibinfo{year}{2014}).
  

\bibitem{Tonolini:2014}
\bibinfo{author}{Tonolini, F.}, \bibinfo{author}{Chan, S.},
  \bibinfo{author}{Agnew, M.}, \bibinfo{author}{Lindsay, A.} \&
  \bibinfo{author}{Leach, J.}
\newblock \bibinfo{title}{{Reconstructing high-dimensional two-photon entangled
  states via compressive sensing}}.
\newblock \emph{\bibinfo{journal}{Sci. Rep.}} \textbf{\bibinfo{volume}{4}},
  \bibinfo{pages}{6542} (\bibinfo{year}{2014}).


\bibitem{Lloyd:2014}
\bibinfo{author}{Lloyd, S.}, \bibinfo{author}{Mohseni, M.} \&
  \bibinfo{author}{Rebentrost, P.}
\newblock \bibinfo{title}{{Quantum principal component analysis}}.
\newblock \emph{\bibinfo{journal}{Nature Phys.}} \textbf{\bibinfo{volume}{10}},
  \bibinfo{pages}{631--633} (\bibinfo{year}{2014}).


  
\bibitem{Lundeen:2011}
\bibinfo{author}{Lundeen, J.~S.}, \bibinfo{author}{Sutherland, B.},
  \bibinfo{author}{Patel, A.}, \bibinfo{author}{Stewart, C.} \&
  \bibinfo{author}{Bamber, C.}
\newblock \bibinfo{title}{{Direct measurement of the quantum wavefunction}}.
\newblock \emph{\bibinfo{journal}{Nature}} \textbf{\bibinfo{volume}{474}},
  \bibinfo{pages}{188--191} (\bibinfo{year}{2012}).

\bibitem{Banaszek:1999}
\bibinfo{author}{Banaszek, K.}, \bibinfo{author}{D'Ariano, G.~M.},
  \bibinfo{author}{Paris, M. G.~A.} \& \bibinfo{author}{Sacchi, M.~F.}
\newblock \bibinfo{title}{{Maximum-likelihood estimation of the density
  matrix}}.
\newblock \emph{\bibinfo{journal}{Phys. Rev. A}} \textbf{\bibinfo{volume}{61}},
  \bibinfo{pages}{10304} (\bibinfo{year}{2000}).


\bibitem{Liu:2012}
\bibinfo{author}{Liu, W.-T.}, \bibinfo{author}{Zhang, T.},
  \bibinfo{author}{Liu, J.-Y.}, \bibinfo{author}{Chen, P.-X.} \&
  \bibinfo{author}{Yuan, J.-M.}
\newblock \bibinfo{title}{{Experimental Quantum State Tomography via Compressed
  Sampling}}.
\newblock \emph{\bibinfo{journal}{Phys. Rev. Lett.}}
  \textbf{\bibinfo{volume}{108}}, \bibinfo{pages}{170403}
  (\bibinfo{year}{2012}).
  
  
\bibitem{Bamber:2014}
\bibinfo{author}{Bamber, C.} \& \bibinfo{author}{Lundeen, J.~S.}
\newblock \bibinfo{title}{{Observing Dirac{\textquoteright}s Classical Phase
  Space Analog to the Quantum State}}.
\newblock \emph{\bibinfo{journal}{Phys. Rev. Lett.}}
  \textbf{\bibinfo{volume}{112}}, \bibinfo{pages}{070405}
  (\bibinfo{year}{2014}).
  


\bibitem{Wu:2013gb}
\bibinfo{author}{Wu, S.}
\newblock \bibinfo{title}{{State tomography via weak measurements}}.
\newblock \emph{\bibinfo{journal}{Sci. Rep.}} \textbf{\bibinfo{volume}{3}}
  (\bibinfo{year}{2013}).
  
    \bibitem{Salvail:2013}
\bibinfo{author}{Salvail, J.~Z.} \emph{et~al.}
\newblock \bibinfo{title}{{Full characterization of polarization states of
  light via direct measurement}}.
\newblock \emph{\bibinfo{journal}{Nature Photon.}}
  \textbf{\bibinfo{volume}{7}}, \bibinfo{pages}{1--6} (\bibinfo{year}{2013}).


\bibitem{Duck:1989}
\bibinfo{author}{Duck, I.~M.}, \bibinfo{author}{Stevenson, P.~M.} \&
  \bibinfo{author}{Sudarshan, E. C.~G.}
\newblock \bibinfo{title}{{The Sense in Which a 'Weak Measurement' of a Spin
  1/2 Particle's Spin Component Yields a Value 100}}.
\newblock \emph{\bibinfo{journal}{Phys. Rev. A}} \textbf{\bibinfo{volume}{40}},
  \bibinfo{pages}{2112--2117} (\bibinfo{year}{1989}).



\bibitem{Malik:2013}
\bibinfo{author}{Malik, M.} \emph{et~al.}
\newblock \bibinfo{title}{{Direct measurement of a 27-dimensional
  orbital-angular-momentum state vector}}.
\newblock \emph{\bibinfo{journal}{Nat. Commun.}} \textbf{\bibinfo{volume}{5}},
  \bibinfo{pages}{3115} (\bibinfo{year}{2014}).
  
 



\bibitem{Osorio:2008}
\bibinfo{author}{Osorio, C.~I.}, \bibinfo{author}{Valencia, A.} \&
  \bibinfo{author}{Torres, J.~P.}
\newblock \bibinfo{title}{{Spatiotemporal correlations in entangled photons
  generated by spontaneous parametric down conversion}}.
\newblock \emph{\bibinfo{journal}{New J. Phys.}} \textbf{\bibinfo{volume}{10}},
  \bibinfo{pages}{113012} (\bibinfo{year}{2008}).
  
\bibitem{Miatto:2011}
\bibinfo{author}{Miatto, F.~M.}, \bibinfo{author}{Yao, A.~M.} \&
  \bibinfo{author}{Barnett, S.~M.}
\newblock \bibinfo{title}{{Full characterization of the quantum spiral
  bandwidth of entangled biphotons}}.
\newblock \emph{\bibinfo{journal}{Phys. Rev. A}} \textbf{\bibinfo{volume}{83}},
  \bibinfo{pages}{033816} (\bibinfo{year}{2011}).

\bibitem{Dada:2011}
\bibinfo{author}{Dada, A.~C.}, \bibinfo{author}{Leach, J.},
  \bibinfo{author}{Buller, G.~S.}, \bibinfo{author}{Padgett, M.~J.} \&
  \bibinfo{author}{Andersson, E.}
\newblock \bibinfo{title}{{Experimental high-dimensional two-photon
  entanglement and violations of generalized Bell inequalities}}.
\newblock \emph{\bibinfo{journal}{Nature Phys.}} \textbf{\bibinfo{volume}{7}},
  \bibinfo{pages}{1--4} (\bibinfo{year}{2011}).


  
\bibitem{Agnew:2011}
\bibinfo{author}{Agnew, M.}, \bibinfo{author}{Leach, J.},
  \bibinfo{author}{McLaren, M.}, \bibinfo{author}{Roux, F.~S.} \&
  \bibinfo{author}{Boyd, R.~W.}
\newblock \bibinfo{title}{{Tomography of the quantum state of photons entangled
  in high dimensions}}.
\newblock \emph{\bibinfo{journal}{Phys. Rev. A}} \textbf{\bibinfo{volume}{84}},
  \bibinfo{pages}{062101} (\bibinfo{year}{2011}).

\bibitem{Leach:2012}
\bibinfo{author}{Leach, J.}, \bibinfo{author}{Bolduc, E.},
  \bibinfo{author}{Gauthier, D.~J.} \& \bibinfo{author}{Boyd, R.~W.}
\newblock \bibinfo{title}{{Secure information capacity of photons entangled in
  many dimensions}}.
\newblock \emph{\bibinfo{journal}{Phys. Rev. A}} \textbf{\bibinfo{volume}{85}},
  \bibinfo{pages}{060304} (\bibinfo{year}{2012}).

\bibitem{Salakhutdinov:2012}
\bibinfo{author}{Salakhutdinov, V.~D.}, \bibinfo{author}{Eliel, E.~R.} \&
  \bibinfo{author}{L{\"o}ffler, W.}
\newblock \bibinfo{title}{{Full-Field Quantum Correlations of Spatially
  Entangled Photons}}.
\newblock \emph{\bibinfo{journal}{Phys. Rev. Lett.}}
  \textbf{\bibinfo{volume}{108}}, \bibinfo{pages}{173604}
  (\bibinfo{year}{2012}).

\bibitem{Tasca:2012}
\bibinfo{author}{Tasca, D.~S.} \emph{et~al.}
\newblock \bibinfo{title}{{Imaging high-dimensional spatial entanglement with a
  camera}}.
\newblock \emph{\bibinfo{journal}{Nat. Commun.}} \textbf{\bibinfo{volume}{3}},
  \bibinfo{pages}{3:984} (\bibinfo{year}{2012}).

\bibitem{Geelen:2013}
\bibinfo{author}{Geelen, D.} \& \bibinfo{author}{L{\"o}ffler, W.}
\newblock \bibinfo{title}{{Walsh modes and radial quantum correlations of
  spatially entangled photons}}.
\newblock \emph{\bibinfo{journal}{Opt. Lett.}} \textbf{\bibinfo{volume}{38}},
  \bibinfo{pages}{4108--4111} (\bibinfo{year}{2013}).



\bibitem{Mosley:2008}
\bibinfo{author}{Mosley, P.~J.}, \bibinfo{author}{Lundeen, J.~S.},
  \bibinfo{author}{Smith, B.~J.} \& \bibinfo{author}{Walmsley, I.~A.}
\newblock \bibinfo{title}{{Conditional preparation of single photons using
  parametric downconversion: a recipe for purity}}.
\newblock \emph{\bibinfo{journal}{New J. Phys.}} \textbf{\bibinfo{volume}{10}},
  \bibinfo{pages}{093011} (\bibinfo{year}{2008}).

\bibitem{Osorio:2013}
\bibinfo{author}{Osorio, C.~I.}, \bibinfo{author}{Sangouard, N.} \&
  \bibinfo{author}{Thew, R.~T.}
\newblock \bibinfo{title}{{On the purity and indistinguishability of
  down-converted photons}}.
\newblock \emph{\bibinfo{journal}{J. Phys. B: At. Mol. Opt. Phys.}}
  \textbf{\bibinfo{volume}{46}}, \bibinfo{pages}{055501}
  (\bibinfo{year}{2013}).

    \bibitem{James:2001}
\bibinfo{author}{James, D. F.~V.}, \bibinfo{author}{Kwiat, P.~G.},
  \bibinfo{author}{Munro, W.~J.} \& \bibinfo{author}{White, A.~G.}
\newblock \bibinfo{title}{{Measurement of qubits}}.
\newblock \emph{\bibinfo{journal}{Phys. Rev. A}} \textbf{\bibinfo{volume}{64}},
  \bibinfo{pages}{052312} (\bibinfo{year}{2001}).
  
\bibitem{Ekert:1995}
\bibinfo{author}{Ekert, A.} \& \bibinfo{author}{Knight, P.~L.}
\newblock \bibinfo{title}{{Entangled quantum systems and the Schmidt
  decomposition}}.
\newblock \emph{\bibinfo{journal}{Am. J. Phys.}} \textbf{\bibinfo{volume}{63}}
  (\bibinfo{year}{1995}).
  
  
  \bibitem{Miller:2013}
\bibinfo{author}{Miller, D. A.~B.}
\newblock \bibinfo{title}{{Self-configuring universal linear optical component
  [Invited]}}.
\newblock \emph{\bibinfo{journal}{Photon. Res.}} \textbf{\bibinfo{volume}{1}},
  \bibinfo{pages}{1} (\bibinfo{year}{2013}).



\bibitem{Bolduc:2013}
\bibinfo{author}{Bolduc, E.}, \bibinfo{author}{Bent, N.},
  \bibinfo{author}{Santamato, E.}, \bibinfo{author}{Karimi, E.} \&
  \bibinfo{author}{Boyd, R.~W.}
\newblock \bibinfo{title}{{Exact solution to simultaneous intensity and phase
  encryption with a single phase-only hologram}}.
\newblock \emph{\bibinfo{journal}{Opt. Lett.}} \textbf{\bibinfo{volume}{38}},
  \bibinfo{pages}{3546--3549} (\bibinfo{year}{2013}).
  
    
  \bibitem{Christandl:2012}
\bibinfo{author}{Christandl, M.} \& \bibinfo{author}{Renner, R.}
\newblock \bibinfo{title}{{Reliable Quantum State Tomography}}.
\newblock \emph{\bibinfo{journal}{Phys. Rev. Lett.}}
  \textbf{\bibinfo{volume}{109}}, \bibinfo{pages}{120403}
  (\bibinfo{year}{2012}).

  
      \bibitem{Berkhout:2010}
  \bibinfo{author}{Berkhout, G.~C.G.},  \bibinfo{author}{ Lavery, Martin P.J.},  \bibinfo{author}{Courtial, J.},  \bibinfo{author}{Beijersbergen, M.~W.} \&  \bibinfo{author}{Padgett, M. J.}
  \newblock \bibinfo{title}{{Efficient sorting of orbital angular momentum states of light}}.
  \newblock \emph{\bibinfo{journal}{Phys. Rev. Lett.}}
    \textbf{\bibinfo{volume}{105}}, \bibinfo{pages}{153601}
    (bibinfo{year}{2010}).
  
  \bibitem{Morizur:2010}
  \bibinfo{author}{Morizur, J.-F.},  \bibinfo{author}{ Nicholls, L.},  \bibinfo{author}{Jian, P.},  \bibinfo{author}{Armstrong, S.}, \bibinfo{author}{ Treps, N.}, \bibinfo{author}{ Hage, B.}, \bibinfo{author}{ Hsu, M.}, \bibinfo{author}{ Bowen, W.}, \bibinfo{author}{ Janousek, J.},    \&  \bibinfo{author}  {Bachor, H.-A.}
  \newblock \bibinfo{title}{{Programmable unitary spatial mode manipulation}}.
  \newblock \emph{\bibinfo{journal}{JOSA A}}
    \textbf{\bibinfo{volume}{27}}, \bibinfo{pages}{2524--2531}
    (\bibinfo{year}{2010}).
    
  \bibitem{Miller:2013}
  \bibinfo{author}{Miller, D.}
  \newblock \bibinfo{title}{{Reconfigurable add-drop multiplexer for spatial modes}}.
  \newblock \emph{\bibinfo{journal}{Opt. Expr.}}
    \textbf{\bibinfo{volume}{21}}, \bibinfo{pages}{20220--20229}
    (bibinfo{year}{2013}).

  
  









\bibitem{Boyd79}
\bibinfo{author}{Boyd, R.~W.}
\newblock \bibinfo{title}{Intuitive explanation of the phase anomaly of focused
  light beams}.
\newblock \emph{\bibinfo{journal}{J. Opt. Soc. Am.}}
  \textbf{\bibinfo{volume}{70}}, \bibinfo{pages}{877--880}
  (\bibinfo{year}{1980}).
  
\end{thebibliography}
\end{document}